\documentclass{llncs}

\usepackage{hyperref}
\usepackage{comment}
\usepackage{dsfont}
\usepackage{stmaryrd}
\usepackage{graphicx}
\usepackage{wrapfig}

\begin{document}

\title{Vulnerability Clustering and \\ other Machine Learning Applications \\
  of Semantic Vulnerability Embeddings}

\author
{
	Mark-Oliver Stehr, Minyoung Kim
}

\institute
{
  SRI International, Menlo Park, CA 94025
}

\date{December 2021}

\maketitle

\pagestyle{plain}

\begin{abstract}
  Cyber-security vulnerabilities are usually published in form of
  short natural language descriptions (e.g., in form of MITRE's CVE
  list) that over time are further manually enriched with labels such
  as those defined by the Common Vulnerability Scoring System
  (CVSS). In the Vulnerability AI (Analytics and Intelligence) project, we investigated different types
  of semantic vulnerability embeddings based on natural language
  processing (NLP) techniques to obtain a concise representation of
  the vulnerability space. We also evaluated their use as a foundation
  for machine learning applications that can support cyber-security
  researchers and analysts in risk assessment and other related
  activities. The particular applications we explored and briefly
  summarize in this report are clustering, classification, and
  visualization, as well as a new logic-based approach to evaluate
  theories about the vulnerability space.
\end{abstract}

\section{Introduction}

For the Vulnerabilities Equities Process (VEP) \cite{vep} and more
generally for Cyber-Security Risk Assessment, it is important to keep
an eye on the big picture of how the cyber-security vulnerability
space is structured and how it is evolving over time. Furthermore, it
should be possible to quickly place newly discovered vulnerabilities
in the space and support subject matter experts with tools to
visualize and navigate the space. In this report on the Vulnerability AI (Analytics and Intelligence)
project, we discuss a fully data-driven approach that leverages
machine learning (especially natural language processing,
classification, and clustering) algorithms to serve as a foundation
for such tools. Our approach is centered around the generation and
evaluation of semantic vulnerability embeddings on which algorithms
for visualization, clustering and classification can be built. It also
serves as a basis for a novel capability for vulnerability space
analysis using logical theories that we briefly explore.

\section{Vulnerability Datasets and Labels}

To employ and validate machine learning algorithms in the proper
application context we have identified the National Vulnerability
Database (NIST NVD) \cite{nvd} as the most suitable dataset. It is fed
from MITRE's CVE list \cite{cve}, and with about $150K$ entries
covering the years $1999-2021$ the most comprehensive list of publicly
disclosed cyber-security vulnerabilities.

From the NVD dataset, we extract the CVE-identifier (which identifies
the vulnerability), the publication date, the natural language
description of the vulnerability. We furthermore extract three types
of labels: The Common Weakness Enumeration (CWE) \cite{cwe} label that
identifies the underlying software weaknesses (typically only one but
sometimes multiple weaknesses are associated with a vulnerability)
that the vulnerability exploits, and a reduced version of the Common
Platform Enumeration (CPE) \cite{cpe} label that identifies the vendor
and product in which the vulnerability was observed. Furthermore, we
extract the Common Vulnerability Scoring System (CVSS) \cite{cvss}
labels related to the most commonly used base metrics, which have
components qualitatively summarizing the impact (e.g. the
confidentiality, integrity, availability impact, and the potential
change of scope) and the exploitability (e.g., attack complexity,
attack vector, privilege/authentication requirement, and need for
user interaction). Quantitative assessments on an ad-hoc scale $0-10$
are also available and sometimes used for our visualizations, but as
these are derived scores (and as we will discuss not reflected in
empirical data), our machine learning algorithms operate directly on
the qualitative features for better precision. We also distinguish
between CVSS V2 and V3 labels,\footnote{To partly bridge the
differences between CVSS V2 and V3, we enrich the V2 labels with
additional components (about user interaction and obtaining
privileges) available in the NVD. The main remaining differences from
V2 to V3 (regarding the base metrics) are the generalization of access
complexity to attack complexity, the introduction of a component
flagging a change of scope (typically related to privileges) and the
consideration of a physical attack vector. A smaller difference is
that confidentiality, integrity, and availability labels (the CIA
properties) use the categories ``None'', ``Low'', or ``High'' in V3 instead of
``None'', ``Partial'', and ``Complete''.} but we will gloss over these
details in this summary.

It is noteworthy that labels have been added by subject matter experts
and are often incomplete and sometimes not very informative (e.g., in
case of CWE labels). Our machine learning workflow is designed to take
labeled and unlabeled data into account.  While our workflow explores
predictors/classifiers for all the label categories mentioned above
and others such as the year/day of publication, we use the
(qualitative) CVSS labels for validating our models and focus on these
in this report.  In the scope of the public dataset available to
us for our analysis, we consider them as the most abstract and most
relevant properties in the context of the VEP and other high-level
risk assessment activities.

\section{Methods and Algorithms}

Our workflow, which is formally defined as a graph with high-level
dataflow dependencies, includes a natural language processing (NLP)
stage applied to vulnerability descriptions and a number of machine
learning models for dimensionality reduction, clustering,
classification, and visualization. The visualization is not only
relevant for the end-user but also allows us to gain a better
understanding of the dataset and potential limitations such as missing
labels.

\begin{figure*}[!htb]
  \begin{center}
	\includegraphics[width=\linewidth]{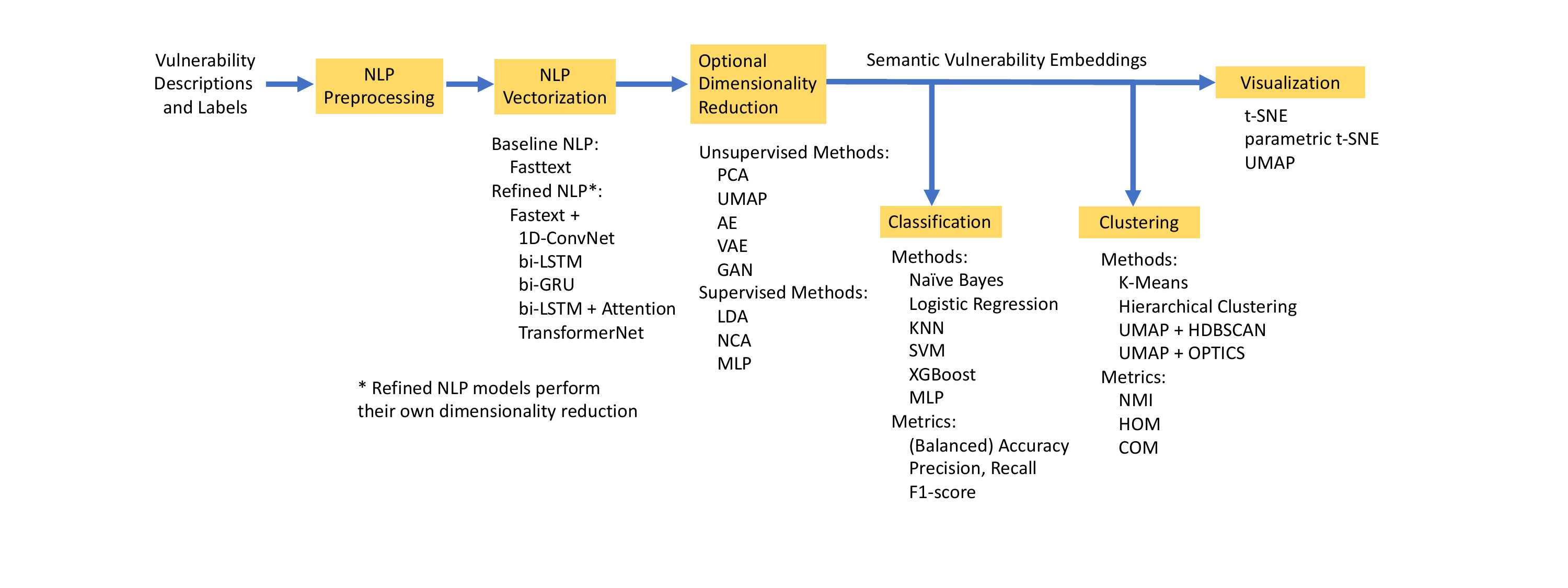}
	\caption{Simplified Excerpt of the Vulnerability AI Workflow to Evaluate
          Semantic Vulnerability Embeddings and Algorithms for Classification, Clustering, and Visualization}
	\vspace{5pt}
	\label{fig-wf1}
  \end{center}
\end{figure*}

A simplified illustration of our basic workflow is shown in
Fig.\ref{fig-wf1}.  An extension of this workflow with more advanced
capabilities will be discussed in Section \ref{sect-palo}. The
workflow was developed and executed using JupyterFlow \cite{Vertes18},
a framework and an engine for interactive and parallel/distributed
machine-learning workflows that we developed in an earlier DARPA
project.

Our NLP stage is based on Fasttext \cite{fasttext1,fasttext2}, which
evolved from earlier work on representing the distributional semantics
of natural language using vector-space embeddings (the most well-known
being Word2Vec, see \cite{wordvectors} for an introduction) that can
be learned by training relatively simple neural networks on a large
corpus of documents. The advantage of Fasttext is that by means of
hashing techniques it can assign useful semantic embeddings to unknown
terms (e.g. misspelled words or filenames containing pieces of natural
language). For this project, we use a Fasttext model trained on Common
Crawl \cite{commoncrawl} which represents a large part of the World
Wide Web and hence also includes many terms and acronyms related to
computer security, hardware/software platforms, and applications.  Our
preprocessing stage uses a number of heuristics to clean up and
normalize the natural language descriptions of vulnerabilities. We
also incorporate composite terms using various sources including a
database of known products/applications. Similar to the Fasttext
approach to sentence embeddings, the abstract vector-space
representation of a vulnerability is the mean of all non-zero
normalized (to the unit norm) embeddings of relevant components (i.e.,
words, acronyms, numbers, filenames, or composite terms that have not
been discarded by our preprocessing) of the description (represented
by 300-dimensional vectors). In the following, we refer to this as the
baseline NLP model that because of its low complexity can be expected
to have the highest generalization capability.  A more refined
representation will play a role in our exploration of neural language
models (in particular, convolutional networks, bidirectional GRU/LSTM
networks, and transformer networks as they can be found in, e.g.,
\cite{nlmsurvey}) that operate on the sequence of vectors representing
the vulnerability description. Some of these refined NLP models are
also equipped with a self-attention mechanism that can take advantage
of the typical structure of CVE vulnerability descriptions to improve
accuracy and other relevant metrics.

Dimensionality reduction techniques operating in the semantic space of
vulnerability descriptions (obtained using our NLP pipeline) include
unsupervised and supervised methods. Unsupervised methods are
Principal Component Analysis (PCA, which is used as a baseline),
Uniform Manifold Approximation and Projection (UMAP) \cite{umap},
Autoencoders (AEs), Variational Autoencoders (VAEs) \cite{vae}, and
Generative Adversarial Networks (GANs) \cite{gan}. As supervised
methods we employed Linear Discriminant Analysis (LDA), Neighborhood
Component Analysis (NCA), and various types of Multi-Layer Perceptrons
(MLPs).

A number of classification methods operating either directly on
semantic embeddings or the reduced representation have been
investigated. They include Naive Bayes Classifiers (baseline),
k-Nearest Neighbor Classifiers, Logistic Regression, Support Vector
Machines (SVM), Gradient-Boosted Decisions Trees (using the XGBoost
\cite{xgboost} framework), and a family of Multi-Layer Perceptrons
parameterized by complexity (number of hidden layers in the range $1
\ldots 3$). The purpose of classifiers in our workflow is
twofold. First, they can be directly used as classifiers by the
end-user, e.g., to label unknown or new vulnerabilities. Classifiers
may also be viewed as a method to extract information from our
semantic representation (e.g., in the form of CPE, CWE, CVSS or other
properties).  Second, and more important to us is that their
evaluation provides us with metrics (e.g. accuracy) to assess the
quality of our vector-space embeddings relative to the
subject-matter-expert labeling. In this way, classifiers serve as an
indirect evaluation tool for previous stages in the workflow, i.e.,
NLP and dimensionality reduction techniques.

Operating in the optionally reduced semantic space, we investigated a
number of clustering algorithms, namely K-Means (baseline),
Hierarchical Clustering (Ward's method), and two different algorithms
for Density-Based Clustering, namely Hierarchical Density-based
Spatial Clustering of Applications with Noise (HDBSCAN) \cite{hdbscan}
and Ordering Points To Identify Cluster Structure (OPTICS)
\cite{optics}. The latter two algorithms can achieve robustness to
noise by using a partial clustering, which means that some
vulnerabilities will not be included in clusters. The selection of
these algorithms is dictated by the need to scale to a large dataset
(in our case we have about 150K vulnerabilities). By operating in the
dimensionality-reduced space scalability can be further improved with
only a minor loss in precision.

\section{Visualization Techniques}

\begin{figure*}[!htb]
  \begin{center}
	\includegraphics[width=\linewidth]{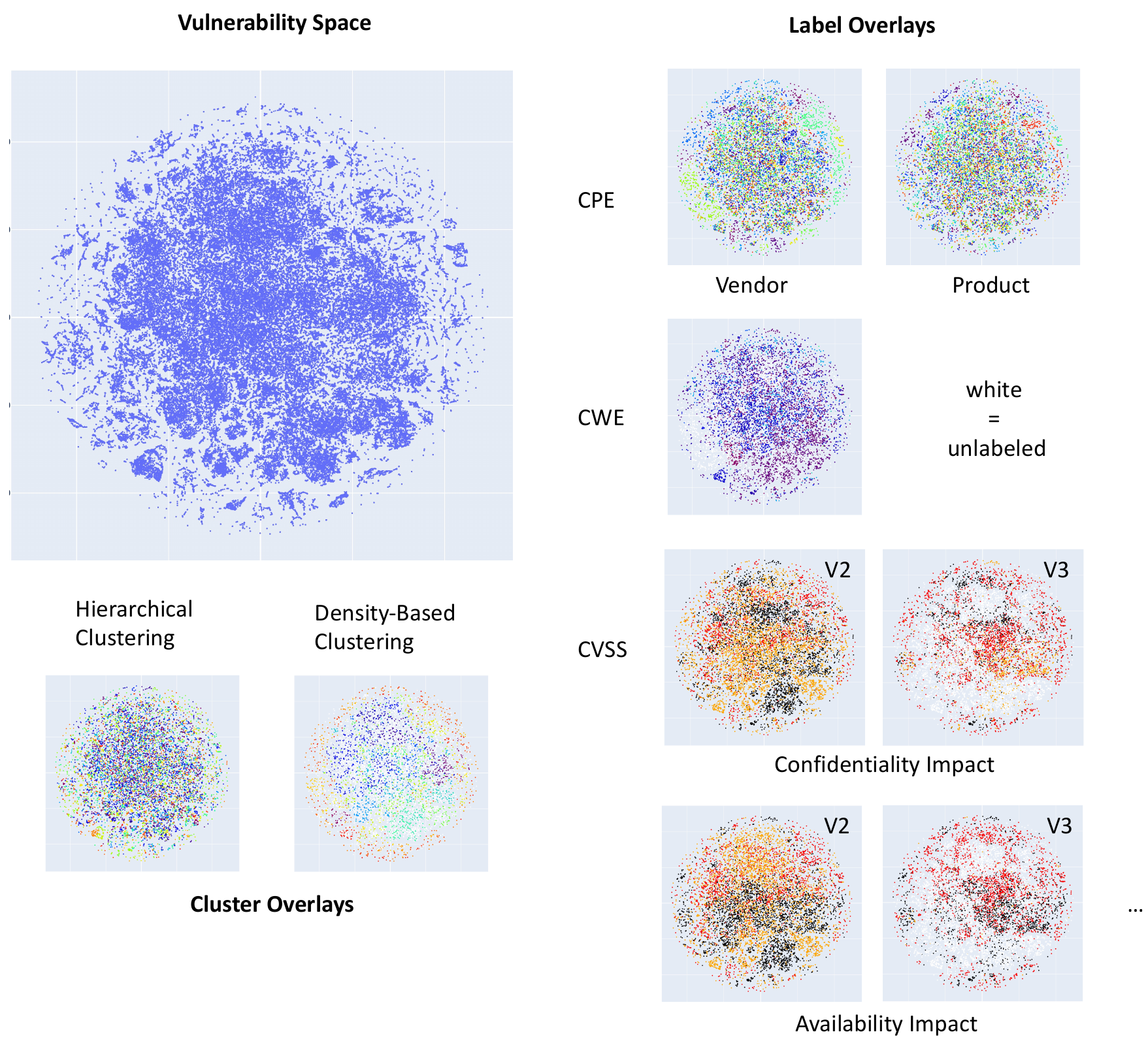}
	\caption{The figure at the top left is a 2D t-SNE projection
          of all semantic vulnerability embeddings (up to 2020, about
          150K) using our autoencoder-based representation. On the
          right, we show a subset of possible overlays using
          color-coded labels. At the bottom, we show clustering
          results as overlays for two algorithms. Clusterings and CWE
          labels are equipped with a hierarchical structure, which is
          captured by color similarity. We use the white color to
          denote absence of labels or no cluster assignment (in case
          of density-based clustering). In case of CVSS labels (we
          only show confidentiality and availability for
          illustration), we use black/orange/red to denote no/low/high
          impact.}
	\vspace{5pt}
	\label{fig-tsne}
  \end{center}
\end{figure*}

\begin{figure*}[!htb]
  \begin{center}
	\includegraphics[width=\linewidth]{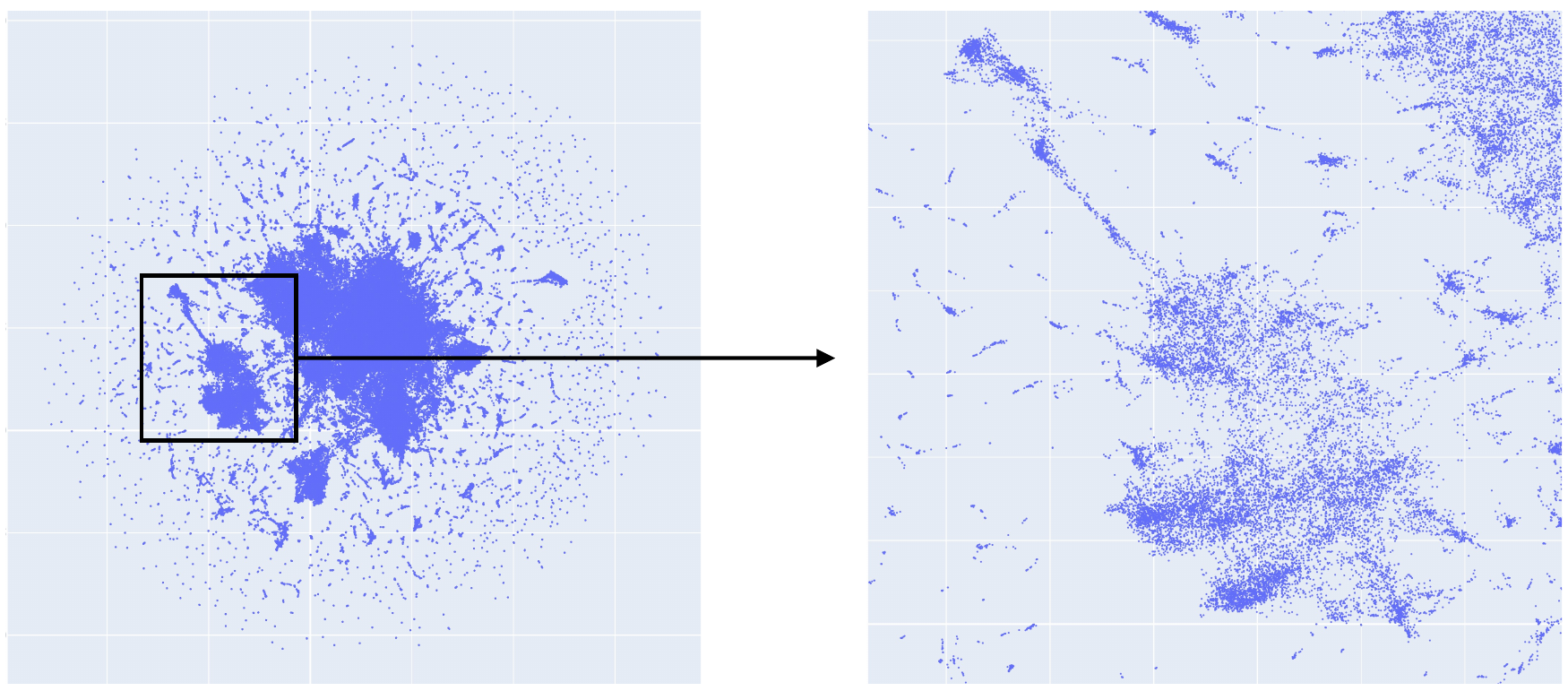}
	\caption{An alternative 2D projection of the same
          vulnerability space as in Fig.~\ref{fig-tsne} using
          UMAP. The large clusters on the left-hand side have
          multiple levels of finer cluster structure that becomes
          visible in the magnification (right-hand side).}
	\vspace{5pt}
	\label{fig-umap}
  \end{center}
\end{figure*}

A number of algorithms for projections, i.e., 2D-visualizations of the
vulnerability space, have been incorporated into our workflow. They
include t-distributed Stochastic Neighbor Embedding (t-SNE)
\cite{tsne}, a version of t-SNE parameterized by neural networks
(parametric t-SNE) \cite{ptsne}, and a version of Uniform Manifold
Approximation and Projection (UMAP) \cite{umap}. All these algorithms
try to map the high-dimensional space into lower dimensions while
approximately preserving the local distribution of the data.  In our
workflow, visualizations are automatically generated for all
combinations of methods for projection, dimensionality reduction, and
clustering. The visualizations (see Fig.~\ref{fig-tsne} and
\ref{fig-umap}) are interactive and hence allow the user to easily
navigate the space of vulnerabilities (a form of semantic
navigation). Additionally, we provide these visualizations for
different subsets, e.g., organized by year, and with color-coded
overlays, e.g., using CVSS, CWE, or CPE labels. It is noteworthy that
unlabelled vulnerabilities are also placed in the same space making it
easy for the user to put them in the context of known and/or
better-understood vulnerabilities.

\begin{figure*}[!htb]
  \begin{center}
	\includegraphics[width=\linewidth]{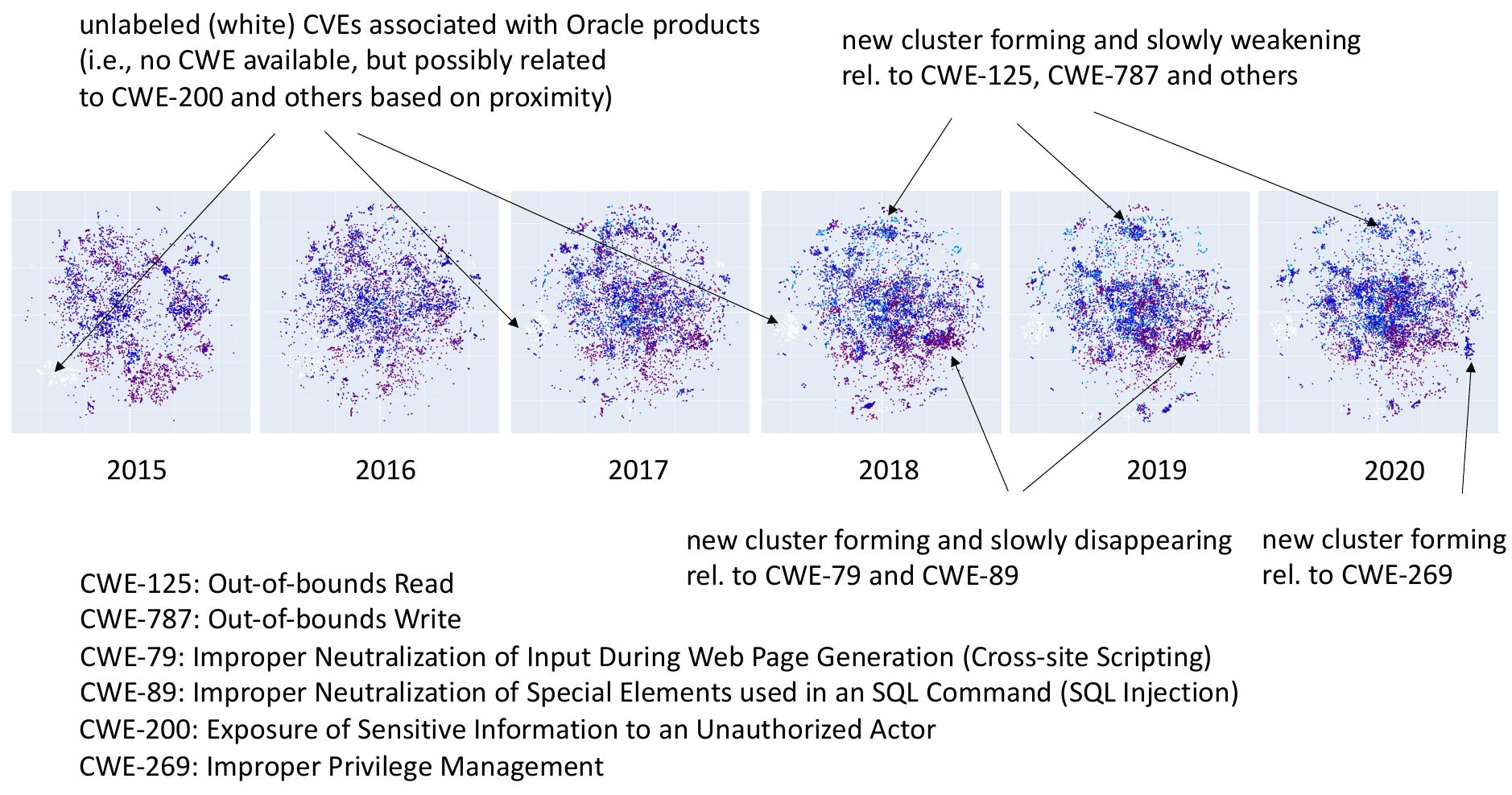}
	\caption{A visualization of the vulnerability space evolving
          over time (with CWE overlay).  We can easily see how the
          overall structure of the space is changing over the entire
          period and how smaller clusters emerge, morph, and sometimes
          disappear over time. Most of the CWEs referenced in this
          figure are still on MITRE's list \cite{top25} of the top 25
          most dangerous software weaknesses for 2021. Many
          vulnerabilities are not equipped with CWE labels (white
          color), but their placement in the space allows us to narrow
          down the likely candidates (even without using
          classification algorithms). Newly discovered vulnerabilities
          can be treated in a similar way.}
	\vspace{5pt}
	\label{fig-evolution}
  \end{center}
\end{figure*}

\begin{figure*}[!htb]
  \begin{center}
	\includegraphics[width=\linewidth]{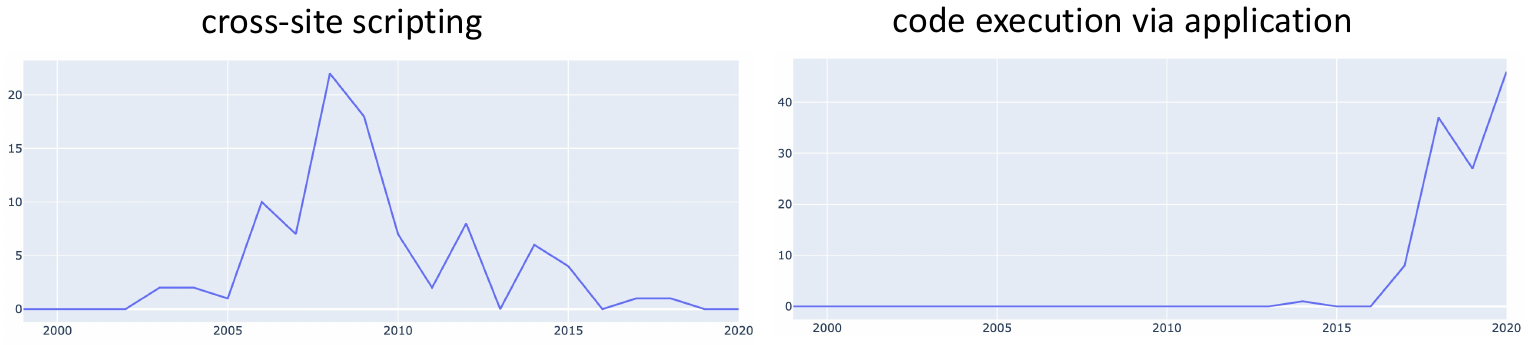}
	\caption{A more fine-grained quantitative trend analysis is
          possible by utilizing the results of clustering algorithms.
          In this figure, we show the number of occurrences of
          vulnerabilities over time in two sample clusters, one
          related to certain types of cross-site scripting
          vulnerabilities and another one related to types of remote
          code execution facilitated by an application. While the
          former class of vulnerabilities seems to have disappeared
          (or became too small to be measured), the latter class can be
          expected to remain a (possibly growing) problem.}
	\vspace{5pt}
	\label{fig-cluster-evolution}
  \end{center}
\end{figure*}

To analyze historical patterns of vulnerabilities and their dynamics
we also implemented a visualization feature that allows us to use
clustering algorithms to depict the evolution of CVE clusters over
time (using annual granularity). This is complementary to our
visualization of the entire vulnerability space over time (see
Fig.~\ref{fig-evolution}), as it allows reducing the complexity by
focusing on single clusters and temporal patterns. To avoid
information overload, we currently use a simple heuristics showing the
top-n largest clusters, but other heuristics taking into account the
trend may be worthwhile to investigate in future work. In general, the
temporal patterns can be quite informative (see
Fig.~\ref{fig-cluster-evolution}), e.g., some vulnerability clusters
lead to a single peak in the number of CVE incidents and then simply
disappear, while some clusters are continuously growing or slowly
declining. Other vulnerability clusters have more complex patterns,
e.g. a smaller peak followed by a much higher peak, which could
indicate that the core problem was not properly extinguished or it
might have been generalized or transferred into other contexts.

\section{Evaluation using Baseline NLP Model}

For the evaluation of algorithms, our base dataset which currently
covers the years $1999-2020$ will be split into a training ($90\%$)
and validation set ($10\%$). The partial data from $2021$ that is
already available in the CVE/NVD database has not been utilized yet
and could serve as an additional validation set in future work. To
partly compensate for concept drift, we suggest that all data should
be incorporated into operational models (periodic retraining of all
stages) once the validation is completed.

We first evaluated the quality of the semantic vulnerability
embeddings generated by the baseline NLP model using classification
algorithms as a validation tool. The main classification metrics we
used are accuracy, balanced accuracy (which corrects for class
imbalance), precision, recall, and the F1-score. Balanced accuracy is
a useful tool to compare the power of the classifier w.r.t. different
types of labels, but less important for the end-user if we assume that
the distribution of new data will be somewhat representative of the
distribution of our dataset. In reality, one would expect some drift
over time, and it would be worthwhile to quantify this is in future
work, and possibly develop techniques (e.g., temporal weighting) to
take it into account, but a challenge would be the additional model
complexity and the possibility of overfitting to very recent data.

The semantic embeddings are either the 300-dimensional vectors from
the baseline NLP model or dimensionality-reduced semantic embeddings
generated by one of the following models. To utilize the Euclidean
norm, all NLP vectors will be normalized to the unit norm before the
reduction is performed.

Our baseline model is Principal Component Analysis (PCA) which has the
advantage of being computationally efficient and as a linear model
maximally preserves the properties of the NLP vector space, e.g., its
compositional structure \cite{word2vec-comp} which often allows us to
interpret addition as a form of feature conjunction.

Another dimensionality-reduction technique is UMAP, which has already
been mentioned as a projection technique for our 2D visualizations.
Here we use it for the reduction of our NLP vectors to smaller
dimensions while approximately preserving the local distribution of
the data (and hence its manifold structure). An advantage of UMAP over
t-SNE is that it is computationally more efficient for
high-dimensional target spaces.

Our Autoencoder (AE) model uses an encoder with two hidden layers (500
and 2000 units) and a symmetrically structured decoder with
relu-activation, dropout-regularization, trained with
mean-squared-error loss using Adam (10000 epochs with batch size
1000).  Our Variational Autoencoder (VAE) model uses a similar
structure but it is a generative probabilistic model with normal
distribution on the latent variables (code space).

Our Generative Adversarial Network (GAN) model uses a slightly more
complex generator based on a normal input distribution (three hidden
layers with 2000, 500, and 500 units) and a discriminator with three
hidden layers (1000 units each). The loss is defined by binary
cross-entropy to distinguish real from fake vulnerabilities (real
vulnerabilities being those from our dataset). Interleaved training of
generator and discriminator is performed as usual (10000 epochs with
batch size 100). To enable the use of our GAN for dimensionality
reduction we also train an inverter (a network with three hidden
layers of 2000, 500, and 2000 units) using Adam with
mean-squared-error loss (for 5000 epochs).

For comparison, we also employed two supervised classical
dimensionality reduction techniques, namely Linear Discriminant
Analysis (LDA) and Neighborhood Component Analysis (NCA). While
performing better than the (unsupervised) PCA, neither of these can
match the performance of our unsupervised autoencoder-based models,
which are highly non-linear in nature, and hence seem more suitable to
capture the complexity of the semantic vulnerability space. As
generative models, VAEs and GANs have their own advantages, which can
be useful for certain applications (e.g., to sample from the
vulnerability space for applications such as those discussed in Section
\ref{sect-palo}), but in our specific workflow aiming at
classification and clustering their performance was inferior even
compared with our baseline PCA.

We also investigated a number of hybrid models combining aspects of
autoencoders (information-bottleneck architecture) and multi-layer
perceptrons (MLP) for supervised dimensionality reduction that we
first applied to Common Weakness Enumeration (CWE) labels.  There are
two primary flavors of MLP models that we refer to as functional and
relational models. The functional model generates a label prediction
as output, whereas the relational model incorporates the label as an
additional input and predicts if it is related to the main input.  The
relational models are clearly more general (and in one variation can
take into account the hierarchical structure of CWEs), but it turned
out that the functional models yield better accuracy, suggesting that
the relational nature of the CVE/CWE associations and their
hierarchical structure does not carry sufficient information to
support a more complex model. We also investigated two orthogonal
flavors of MLPs (classification and regression MLPs, where the
regression model tries to be more semantically accurate by mapping
CWEs into a semantic space induced by their natural language
descriptions). Again, it turned out that the simpler
classification-based model gave superior performance.

Since our objective is to amplify VEP-relevant properties, we next
applied the functional and classification-based MLP (simply referred
to as MLP in the following) to CVSS labels (specifically V2) to obtain
a dimensionality-reduced representation of all CVEs with good
performance (of the subsequent workflow) not only on V2 but also on
V3-labels (an example of transfer learning). This is quite fortunate
because V3 labels are only available for the last few years and
expected to be even more prevalent in the future. Our hypothesis that
the CWE labels could be used for transfer learning and hence
improve the classification performance for CVSS labels was not
confirmed, most likely to the quality and limited information content
of CWE annotations (a few general classes of CWEs dominate the
dataset, despite of the elaborate and fine-grained CWE taxonomy).

To compare the relative performance of the dimensionality reduction
techniques we use relatively simple classifiers including a family of
MLPs (with up to 3 hidden layers), but also a number of classical ML
algorithms, in particular, Naive Bayes, Logistic Regression,
$k$-Nearest Neighbor Classifiers (varying $k$ between $1$ and $100$),
Support Vector Machines (using RBF kernels) and Gradient-Boosted
Decisions Trees (using XGBoost \cite{xgboost}). While the classical
algorithms cannot achieve the performance of MLP classifiers in this
application, they are in most cases simpler models and have their own
advantages (e.g. in terms of explainability). In our relative
performance evaluation, they turned out to be useful tools to better
understand the quality/complexity of our embeddings.  For example, the
semantic vulnerability representation (reduced or not) seems to be too
complex to be interpreted by linear models and SVMs with useful
results.

In summary, our evaluation shows that autoencoder-based
representations (in particular, the Autoencoder Model) yield superior
performance in terms of accuracy among all unsupervised dimensionality
reduction techniques, while the MLP-based model shows the best
performance for classification and supervised dimensionality
reduction.

After the NLP stage, (optional) dimensionality reduction, and
(optional) classification, the next stage in our workflow is
clustering which is inherently unsupervised and requires another set
of metrics. To properly incorporate the distribution of the dataset,
information-theoretic metrics should be used \cite{vmeasure}. A common
metric is Normalized Mutual Information (NMI), which is symmetric and
measures the relationship (degree of coincidence) between predicted
clusters and clusters generated by ground truth (in our case
approximated by labels).  Two other import information-theoretic
metrics are homogeneity (HOM) and completeness (COM). The former
measures how uniform the predicted clusters are (in the sense that
ideally all elements should have the same label), whereas the latter
measures how well the labels are represented by the predicted clusters
(ideally, elements with the same label should be assigned to the same
cluster).  For many reasons, most importantly the fact that all
clustering algorithms are based on heuristics to achieve computational
feasibility and are parameterized to bias/constrain their results,
ideal conditions are typically not achievable (see also
\cite{clustering-impossible}), so that our main objective with these
metrics is to compare the performance of different semantic
representations and clustering algorithms relative to each other.  In
our analysis, algorithms are typically parameterized to generate
fine-grained clusters (several thousand) to provide a high degree of
detail. Hence, homogeneity plays the most important role and is
increased at the expense of completeness.  As a summary measure, we
use NMI with normalization by arithmetic averaging. It is identical to
the V-measure \cite{vmeasure} (or validity measure), which is defined
as the harmonic mean of homogeneity and completeness, quite similar to
the F1-score, which is the harmonic mean of precision and recall. It
has been shown to have many properties that make it a useful (albeit
not perfect) measure to compare algorithms independent of the size of
the data set and the number/size of clusters.

These and other classical metrics are applied to the evaluation of the
four clustering algorithms employed in our workflow. First, we have
K-Means, which is simple and intuitive but severely limits the shape
to clusters (convexity assumption). Then there is Hierarchical
Clustering, which performs bottom-up agglomerative clustering by
recursively merging pairs of clusters starting from singletons using
simple heuristics (we use Ward's method to take into account
variance).  The method is quite intuitive and has the additional
advantage that as a by-product a meaningful hierarchy is generated.
Finally, there is the class of density-based algorithms of which we
employed HDBSCAN \cite{hdbscan} and OPTICS \cite{optics}. Both
identify clusters based on high-density regions (without convexity
limitations) of the data distribution but use different heuristics to
determine what constitutes a cluster. Both algorithms can deal with
noise by avoiding the creation of new clusters unless a certain
minimum cluster size is reached. To compensate for unsatisfactory
performance in high-dimensional spaces (even with only 10 or 20
dimensions), we combine the density-based algorithms with an
additional dimensionality-reduction step which uses UMAP to reduce the
dimension by a factor of two before the clustering algorithm is
applied.

Our analysis based on CVSS labels (V2 and V3) shows that while
Hierarchical Clustering shows performance similar to K-Means in our
application, the improvement by density-based algorithms in all
information-theoretic metrics (NMI, HOM, COM) clearly stands out
across many comparable parameter settings and underlying models.  We
hypothesize that this at least is partly due to their capability to
deal with noise, which in our context is a combination of noise in the
natural language descriptions and labels. Another contributing
factor may be that density-based algorithms can more reliably identify
lower-dimensional manifolds of arbitrary shapes that are created by
similar vulnerabilities. We also observed that OPTICS shows
consistently better performance (NMI, HOM, COM) than HDBSCAN for most
models in our baseline NLP workflow (supervised and
unsupervised).\footnote{Interestingly, this is also the case of our
MLP-based representation but not the case for our transformer-based
models (see next section), where OPTICS scores better than HDBSCAN on
HOM but worse on COM and NMI. This indicates a different tradeoff that
might be caused by higher bias baked into the transformer-based
representations. In fact, a closer examination shows that this is the
case for both of our attention-based models, but not for the other
refined NLP models.}

For dimensionality reduction recall that our NLP representation of
vulnerabilities has 300 dimensions, but to obtain models with high
generalization potential (and as a way to increase robustness to
concept drift) it is desirable to reduce the dimension as much as
possible as long performance does not suffer substantially.  We
experimented with 10 and 20 dimensions and found that 20 dimensions
result in notably better performance than 10, but higher dimensions do
not yield improvements that are worthwhile the additional model
complexity (which also affects runtime). Sample performance results
for MLP classifiers and two clustering algorithms can be found in the
appendix (Figs.\ref{fig-classifier-eval} and \ref{fig-clustering-eval}).

\section{Evaluation using Refined NLP Models}

Up to this point, the evaluation of our algorithms was based on
300-dimensional vector-space embeddings generated by the NLP stage of
our workflow with an optional dimensionality reduction to 10 or 20
dimensions. Recall, that in our baseline NLP model, the vector-space
embeddings were defined simply as a mean of all non-zero normalized
embeddings of the relevant components (i.e., words, acronyms, numbers,
filenames, or composite terms) of the vulnerability description. This
means that we have effectively employed a multiset abstraction of
vulnerability descriptions which cannot take into account their
sequential structure, e.g., natural language grammar or the typical
organization of the vulnerability descriptions, which usually consist
of multiple sentences providing an increasing level of detail.  In this
section, we discuss an extension of our workflow with a number of
supervised NLP models that directly operate on the sequence of vectors
corresponding to the components of the vulnerability description with
the idea to learn and exploit the high-level structure of these
descriptions in our dataset. It was not a priori clear if better
performance (that can actually generalize) can be achieved due to the
higher model complexity, but our evaluation shows that this is indeed
the case.

Indeed, as the precision of the vulnerability clusters depends on the
underlying NLP model, it makes sense to advance our baseline model by
taking into account more information from the available labels in the
dataset. To this end, we studied a number of supervised NLP models, in
particular convolutional networks and bidirectional LSTMs (Long
Short-Term Memory) and bidirectional GRU (Gated Recurrent Unit)
networks, which have found many applications in text classification
(see, e.g., \cite{nlmsurvey} for a survey covering these models).  In
addition, we investigated neural network models utilizing a
self-attention mechanism to focus on particular parts on the
vulnerability description in a context-dependent manner. In
particular, we implemented a version of bidirectional LSTMs with a
simple notion of attention (similar to \cite{han}) and a simplified
version of the more recently developed transformer model
\cite{transformer} for NLP that in our setting will be applied to a
classification problem rather than a transformation/translation
problem that the more general model is targeting.

In summary, the following NLP models, all based on neural networks,
were implemented and evaluated: 1D-convolutional networks,
bidirectional LSTM networks (single layer of LSTM with global
pooling), bidirectional GRU networks (single layer of GRU with global
pooling), bidirectional LSTMs with simple dot-product attention, and
transformer networks (simplified version suitable for
classification). The main features of the latter are the more powerful
notion of multi-head attention and the use of positional text
encodings (which eliminate the need for recurrent or convolutional
structure). All models were trained as classifiers with CWE and CVSS
V2 labels, and are designed to perform their own dimensionality
reduction (information-bottleneck architecture).

All these models are then employed for supervised dimensionality
reduction (in the context of our larger workflow), and we found that
the accuracy of CVSS classifiers and the quality of clusters (based on
information-theoretic metrics) is indeed further improving. We focused
on the NLP models supervised using CVSS labels, as these are the most
abstract properties that are also used for cluster validation. Our
analysis shows that the transformer model (without fine-tuning) shows
the best tradeoff between generalization capability and classification
accuracy (and related information-theoretic metrics for
clustering). It also yields better performance than our MLP-based
model (the best supervised model built on our baseline NLP).  We also
evaluated variations of all models to retrain (fine-tune) word
embeddings, which naturally leads to much higher model complexity and
has not resulted in additional performance improvements.

The autoencoder model (for our baseline NLP) shows lower performance
but it still remains relevant. As a fully unsupervised model, it is
not biased by the subject-matter-expert CVSS labels and hence might
have advantages for certain applications. In addition, we hypothesize
that due to its lower complexity it may be less prone to concept
drift, which is a separate issue that would be worthwhile to
investigate in the future.  In summary, we consider both the
unsupervised autoencoder-based model as well as the supervised
transformer-based model as the best-performing models arising from our
investigation.  For a comparison of sample performance results for
these two types of models, we again refer to the appendix
(Figs.\ref{fig-classifier-eval} and \ref{fig-clustering-eval}).

\section{Evaluating Theories about the Vulnerability Space}\label{sect-palo}

\begin{figure*}[!htb]
  \begin{center}
	\includegraphics[width=\linewidth]{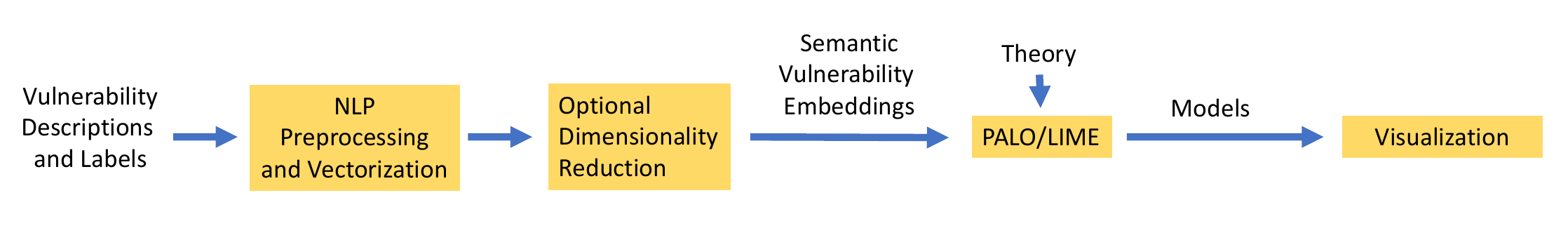}
	\caption{Extension of the Vulnerability AI Workflow with PALO/LIME for Evaluating Theories about the Vulnerability Space}
	\vspace{5pt}
	\label{fig-wf2}
  \end{center}
\end{figure*}

In the following section, we discuss another application of our
semantic vulnerability representations, that has the potential to
further enlarge the toolbox for security risk assessment. Inspired by
discussions with the Vulnerability AI team members, we implemented an
experimental capability to test arbitrary logical theories in the
context of our vulnerability dataset.  It utilizes SRI's Probabilistic
Approximate Logic (PALO) framework to integrate logic and machine
learning \cite{palolime}. PALO was developed in a previous DARPA
project in a biological context (network modeling)
\cite{palolime-rta}, and is now for the first time being applied in
the security domain.  Its implementation in SRI's Logical Imagination
Engine (LIME) uses a general form of Bayesian inference to synthesize
probabilistic models under logical constraints (representing, e.g.,
domain knowledge or hypotheses). This tool may allow us to shed some
light on certain hypotheses about the structure of the vulnerability
space (e.g., regarding compositionality, see below) and more generally
provide a (probabilistic) relational and hence graph-theoretic
perspective with new types of visualizations and new ways to look at
interesting subspaces. The experimental integration of PALO/LIME into
the Vulnerability AI workflow is illustrated in Fig.~\ref{fig-wf2}.

For the Vulnerability AI project, we added support to deal with very large
relations over the vulnerability space and visualize them partially as
graphs (using graphml as a description language) with informative
labels making them easy to navigate for the user.  As a first test
case, we applied our tool to a logical theory of compositionality that
tries to exploit the (partially) compositional structure of the
underlying NLP models \cite{word2vec-comp}.  The specific question
(that was posed in a meeting with Galois) is if our NLP-based models
have learned a compositional structure that can be in any way related
to (or some approximation of) the composition of vulnerabilities in
the real world as it plays a role in vulnerability chaining.  To
simplify the problem (and the formalization) we focused on a binary
notion of compositionality that is reflexive and symmetric (i.e.,
formally a similarity relation) but usually not transitive.
Reflexivity in our probabilistic approximately logic means that
similar vulnerabilities are considered composable, the justification
being that even a single vulnerability could be used multiple times in
a single attack (self-composability), but many variations of this
axiomatization are clearly possible and could be explored in the
future. One of the axioms in our theory relates this hypothesized
notion of vulnerability composability to composability in the
NLP-sense, that for simplicity we approximate by the mean of the
semantic vulnerability vectors, which has a very intuitive
representation in the vulnerability space. More details about the
theory can be found in Fig.~\ref{fig-theory}.

We then conducted a systematic exploration of the models of our
vulnerability composition theory. To this end, we selected three
models as base models for our analysis, namely Principal Component
Analysis (PCA), Autoencoders (AE), and transformer networks. The
PCA-based models are of interest here, because as a linear
transformation PCA will maximally preserve the compositional structure
of the NLP space, while the other models can be expected to distort
that structure due to their non-linear nature. Nevertheless, a visual
inspection shows that all three base models lead to interesting graphs
that focus on different subsets of the larger vulnerability space. In
our preliminary analysis of these graphs (see
Fig.~\ref{fig-theory-graph} for a sample) we observed that highly
connected (and hence in our theory highly composable) vulnerabilities
are often associated with features such as "remote code execution"
that can be critical steps of larger vulnerability chains. The nodes
in their neighborhood often represent vulnerabilities that arise in a
similar context and hence may be candidates for compositions, but on
the other hand the information in the vulnerability descriptions is
not sufficient to verify composability.  Hence, the graphs resulting
from this basic theory must be interpreted with a grain of salt and
can at best be seen as an idea generator that requires additional
filtering by a subject matter expert.

With about 150K nodes, the full graph of the vulnerability space is
expected to be huge and impossible to visualize, but the inference and
computation time is reasonable (several hours per model utilizing a
state-of-the-art GPU) and the resulting small subgraphs (with up to
1000 nodes) are easy to navigate and to interpret thanks to the use of
automatic layout algorithms and graph analysis tools for
visualization. While these preliminary results show that our
experimental tool scales up to the high complexity needed to explore
theories about the vulnerability space, it is not clear if the
particular logical theory based (only) on NLP compositionality is
general/restrictive enough to be of practical interest. Hence, it
would be worthwhile to consider alternative theories of composability
in future work, e.g., by incorporating axioms about CWE, CVSS, or CPE
labels from our dataset or by incorporating other datasets, features,
or risk metrics into the theory. Another pragmatic approach is to
develop an inductive theory of composability. It could start with
empirical data about observed vulnerability chains (which is still
needs to be collected) and utilize semantic vulnerability embeddings
(and related similarity or equivalence relations) to extend this
notion to new vulnerabilities and in this way derive new likely
candidates for compositions.

Finally, we should remind ourselves that the basic compositionality
theory only served as a test case for our experimental tool. We expect
that the tool can be applied to a wide range of relational and
quantitative theories that test other hypotheses or highlight hidden
structures/patterns in the vulnerability space that are not apparent
in our more abstract cloud-based or cluster-based visualizations.

\section{Related Work}

The core of our work is concerned with the unsupervised learning of
semantic vulnerability embeddings, which can serve as a basis for
visualization, clustering, and classification models for quite
different types of applications. We have also investigated the use of
supervised methods to learn vulnerability embeddings that are
naturally biased by the training labels (in this report we focused on
CVSS labels but we also explored the use of CWE labels). Not not
surprisingly the latter models yield better performance for tasks
related to those labels, which will be useful, e.g., to improve
clustering performance in an application (e.g., in the context of the
VEP) where high-level properties such as those related to CVSS labels
play an important role.

In the following, we discuss some related work that utilizes NLP
techniques to extract information from vulnerability descriptions.  In
particular, the named entity recognition/extraction (NER) problem has
attracted considerable interest in the cyber-security research given
the large number of unstructured textual information sources that are
relevant in that domain. While our work focusses on the representation
of vulnerabilities and their classification (which is also a form of
information extraction), there is an overlap with NER and similarity
in the types of models employed that we briefly discuss using a few
very recent works.  For a more complete review, we refer to
\cite{ner-review}.

For example, the top-level objective of \cite{attackmodel} is to
translate vulnerability descriptions from the NVD into interaction
rules for MulVAL (a network security analyzer based on Datalog). The
main problem in this context is attack-entity extraction (a form of
NER), i.e. the extraction of specific attack-related entities
(operating system, attack vector, attack technique, impact, etc.)
from the vulnerability description.  To this end, the authors use
Word2Vec (a predecessor of Fasttext). Instead of using a pretrained
model, they specifically train a Word2Vec model on the NVD
descriptions, and show that its performance is superior to pretrained
ELMo and BERT models for this particular task, which is not too
surprising, but might also mean that the Word2Vec model is fit to the
particular dataset which is quite small for training a language
model. Clusters of Word2Vec embeddings typically contain instances
belonging to the same entity (e.g., operating system).  Attack-entity
extraction is then performed by feeding the word embeddings for the
description into a bidirectional LSTM classifier.  To enable
supervised learning, the vulnerability descriptions were labeled
manually by a large number of students.  Different from our work,
embeddings are only used at the level of words rather than for
vulnerabilities, and it is not clear how well the models would
generalize to new vulnerabilities/descriptions (which might be less
important for this particular application). On the other hand, some
aspects of this approach (e.g., the need to complete missing
information) might actually benefit from clustering at the level of
vulnerabilities, that is by taking into account the bigger picture
that we focus on in our work.

In other recent work \cite{bilstm-crf-ext}, the authors are exploring
general models for NER in the cyber-security domain. As a starting
point, they use the popular bidirectional LSTM model with a
Conditional Random Field layer (bi-LSTM-CRF) \cite{bilstm-crf} and
investigate two extensions, multi-head attention and the use of
domain-specific dictionaries (providing additional input to the
network to improve performance for rare entities).  The advantages of
using an attention mechanism are clearly demonstrated by improved
precision (which is independent of the improvement due to
dictionaries). We also observed improvements by using dot-product
attention in the context of simpler bidirectional LSTM classifiers,
but finally identified the transformer-based classification model that
is inherently based on attention (and in this way completely avoids
the need for recurrent networks) as providing the best performance
tradeoffs for our application. An interesting question remains how
well our semantic vulnerability representation that is not trained for
NER would perform for related classification tasks and if it could
provide any other benefits (e.g., due to the highly compressed semantic
representation).

Another line of work is the use of NLP techniques to better quantify
exploitability risk, and hence address one of the shortcomings of
CVSS.  The CVSS base metrics provide only a very limited static
picture, and a dynamic data-driven approach to assess exploitability
is clearly more appropriate for the VEP and other risk assessment
activities. While CVSS temporal metrics have been defined to address
this problem, their adoption seems to be much more limited.
Exploitability was already investigated in the early work
\cite{exploit-svm}, where an SVM was trained to predict exploitability
based on the very limited data sources available at that time (CVE and
OSVDB, using data up to 2007). This work nicely illustrates the large
discrepancy between predicted exploitability (based on empirical data)
and the CVSS exploitability score, which is another reason why it is
not used in our work.  A more recent promising approach is to build
parsimonious probabilistic models of exploitability as in \cite{epss},
which introduced the Exploit Prediction Scoring System (EPSS). In
addition to features based on keyphrases extracted from CVE/NVD
descriptions (a simple form of NLP), the authors incorporate various
exploit databases (covering proof-of-concept and weaponized exploits)
to determine the probability that a vulnerability will be exploited
after publication and propose regularized logistic regression (Elastic
Net Model) as a predictive model, which seems quite appropriate given
the still relatively limited amount of data available about exploits.
Utilizing a broader set of additional data sources, the recent work
\cite{expexploit} addresses a similar problem with a notion of
expected exploitability and develops neural-network-based models to
predict exploitability with particular focus on addressing label
noise. All these approaches (which use simple forms of NLP not based
on semantic embeddings) suggest that our semantic vulnerability
representations (possibly refined by utilizing additional data sources)
could serve as a unifying basis for similar approaches and offer
potential improvements.  Indeed, as stated in \cite{expexploit} this
direction has been briefly explored but seems challenging and is left
for future work: ``Overall, our result reveals that creating higher
level, semantic, NLP features for exploit prediction is a challenging
problem, and requires solutions beyond using off-the-shelf tools. We
leave this problem to future work.''

\section{Conclusions and Directions for Future Work}

For the Vulnerabilities Equities Process (VEP) a big picture view of
the vulnerability space can be immensely useful. The big picture
allows the expert to understand the structure of the space and how it
is evolving over time at an aggregate level and through the use of
clustering techniques recognize important trends that may influence
the risk assessment of new vulnerabilities that are being
discovered. On a smaller scale, newly discovered vulnerabilities can
be put into context of existing vulnerabilities simply by allowing the
user to navigate the semantic vulnerability space. Classification
algorithms such as those investigated in our work can exploit 
vector-space embeddings to label newly discovered vulnerabilities, but
may also be used to label existing vulnerabilities in the dataset that
have not been labeled due to limited resources. Classifiers and
visualizations of the semantic space may also be used to find problems
with the existing annotations, e.g., mislabeling or other
inconsistencies.

In addition to the informal use of our semantic vector-space
embeddings and their use as part of classification and clustering
algorithms, there are other interesting applications.  One application
that we briefly explored (and which can be further generalized as
discussed below) is the representation of vulnerabilities and their
relations in the context of theories that need to be tested. Other
applications that we did not investigate include the support of
semantic search algorithms to find similar vulnerabilities and the
pruning of a search space of vulnerabilities in the context of
symbolic formal methods, e.g., to check or refute certain logical
properties. Semantic vulnerability embeddings and their associated
notion of similarity (or equivalence in case of clustering) can also
be used to extend empirical data about vulnerabilities to anticipate
new risks.  A security analyst, for example, may utilize our
embeddings and clusterings to explore and anticipate candidates for
new vulnerability chains based on existing ones.

In the remainder of this section, we will briefly discuss directions
for possible future work. First of all, there is simply the further
evolution of our models by integrating with additional data sources.
One limitation of the CVE/NVD dataset is the small amount of detail
about the vulnerability represented in the description. Additional
information may come from evaluating security blogs or even source
code repositories that are partly referenced in the CVE/NVD
entries. To process larger security-related documents, enhanced NLP
models (including NER, as discussed under related work) and
domain-specific training (on a large corpus, e.g., blogs, research
papers) would be worthwhile to explore.

Another dimension that would be important to add to our models is
empirical risk, which ideally should be regarded as a dynamic and
context-dependent notion. One key aspect, empirical exploitability, is
tracked by a number of commercial efforts, and we have already
discussed the work on the Exploit Prediction Scoring System
\cite{epss} and on Expected Exploitability \cite{expexploit}, two
notions that could be potentially predicted more accurately by
utilizing semantic vulnerability embeddings (ideally enriched by
taking into account commercial data sources) as a unifying foundation.

A remaining limitation of our current machine learning workflow is
that we are focusing on the lowest level of the MITRE cyber-security
abstraction hierarchy where the most amount of data is available,
namely at the level of specific vulnerabilities, i.e., MITRE's Common
Vulnerability Enumeration (CVE), but for many other applications such
as monitoring, forensics, and defense, the larger context of
vulnerabilities is important, which would require taking into account
higher levels of abstraction, e.g., the Common Weakness Enumeration
(CWE) \cite{cwe} (which we already used in some of our classifiers),
the Common Attack Pattern Enumeration and Classification (CAPEC)
\cite{capec}, and the Framework for Adversarial Tactics, Techniques \&
Common Knowledge (ATT\&CK) \cite{attack}. While moving to higher
levels of abstraction is very challenging due to the limited amount of
empirical data, new modeling techniques that integrate
state-of-the-art machine learning with logical (in particular
relational) modeling may help to overcome the current limitations
through the use of formal theories (as a representation of domain
knowledge or hypotheses) as a natural counterpart to the input data.

One possible approach to address these issues is to further generalize
our work to incorporate logical theories based on Probabilistic
Approximate Logic (PALO) \cite{palolime} that we already applied to
test a theory of vulnerability composition.  This approach has been
originally developed in the DARPA Rapid Threat Assessment project
\cite{palolime-rta} and our preliminary experiments show that it is
applicable to our vulnerability dataset as well. Based on these
foundations, a natural future direction is to develop a tool that
allows us to bridge multiple levels of abstractions and test virtually
arbitrary logical theories and hypotheses related to cyber-security
risks by combining multiple sources of information. Such a tool should
also have the capability to generate relational/graphical models as a
new way of visualizing/explaining different subsets of vulnerabilities
in the context of other entities and relations referenced by flexible
user-defined theories.

\vspace*{1cm}

\noindent
{\bf Acknowledgments} We gratefully acknowledge contributions from the
entire Vulnerability AI team, in particular Pat Lincoln, Steven Cheung, and
Vinod Yegneswaran for the idea to explore vulnerability clustering,
for helping to select the proper datasets, and for pointers to related
work.  We are also grateful for feedback from Galois, in particular
John Launchbury for the idea to investigate the relationship between
vulnerability composition and composition in vector-space NLP
models. Finally, we would like to thank Carolyn
Talcott for contributing many ideas, especially in the context of
PALO/LIME and the overall machine learning workflow.

\clearpage

\appendix

\section{Other Visualizations and Results for Selected Models}

\begin{figure*}[!htb]
  \begin{center}
	\includegraphics[width=\linewidth]{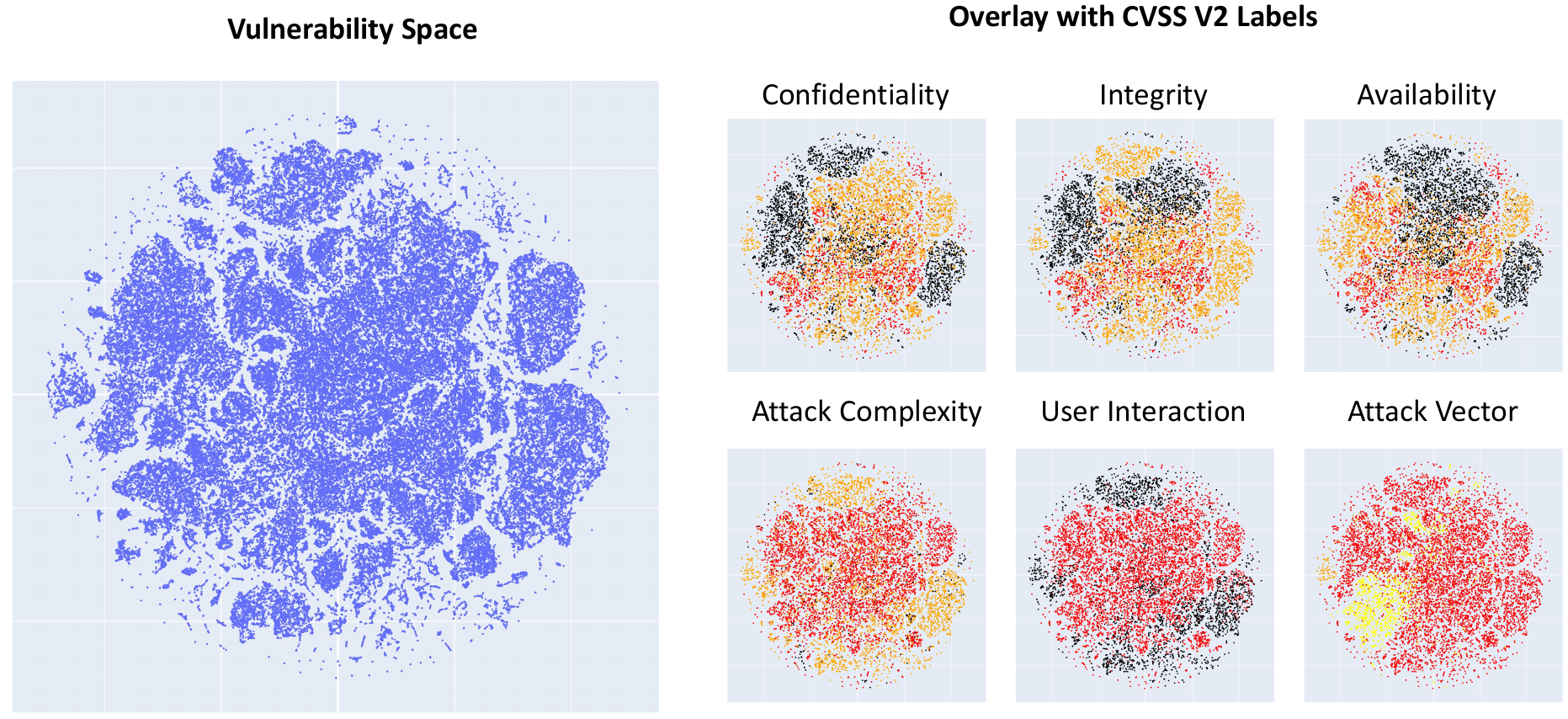}
	\caption{On the left, we see a 2D t-SNE projection of the vulnerability space using
          transformer-based embeddings. It is instructive to compare it with
          the representation based on the autoencoder model in Fig.~\ref{fig-tsne}.
          The supervised nature of the model leads to a visibly more pronounced cluster structure and
          a slightly better fit with the CVSS labels (shown on the right-hand side using a subset of the CVSS V2 labels).
          We use red to denote higher risk categories, orange for medium risk, followed by yellow and black for lower or no risk.
          In the case of user interaction, red means it is not required. In the case of the attack vector,
          red, orange, and yellow means network, adjacent network, and local network, respectively.}
	\vspace{5pt}
	\label{fig-tsne-trans}
  \end{center}
\end{figure*}

\clearpage

\begin{figure*}[!htb]
  \begin{center}
	\includegraphics[width=0.8\linewidth]{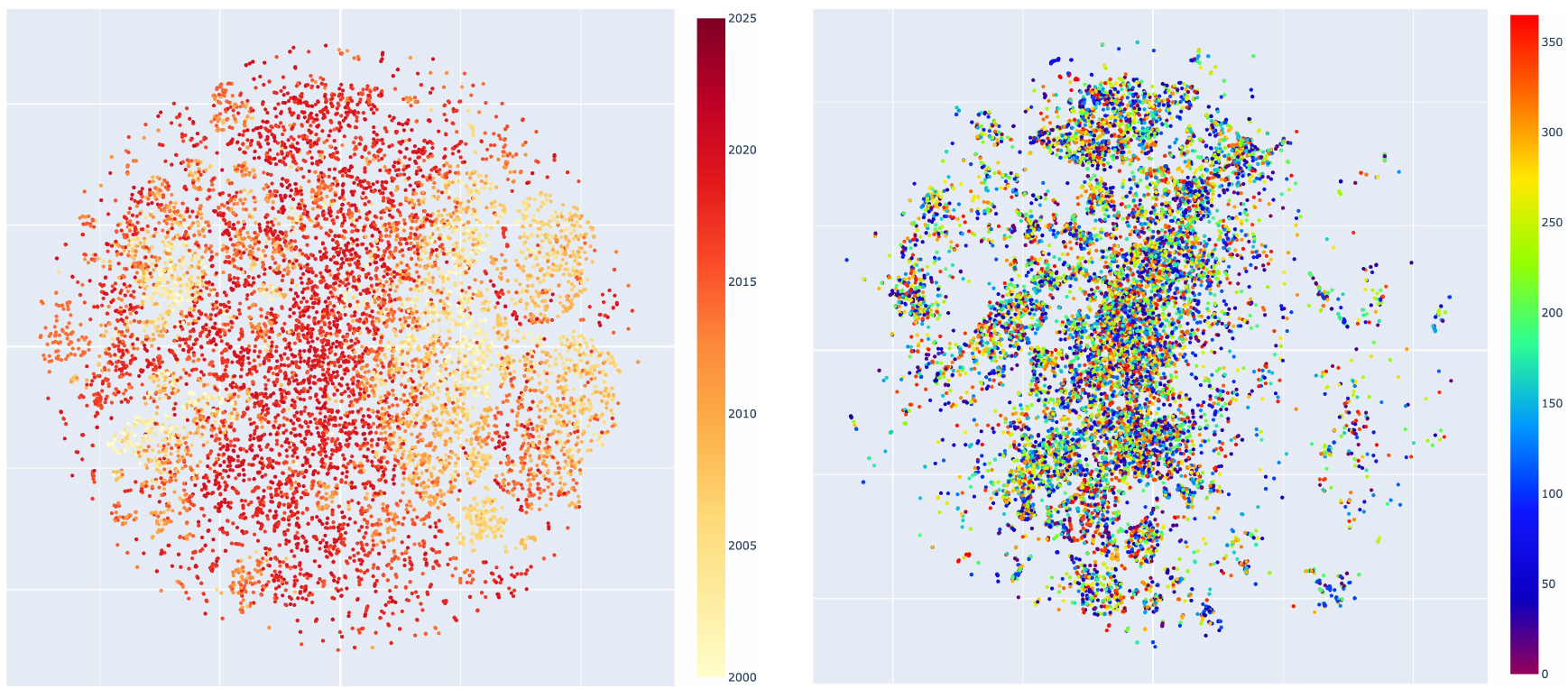}
	\caption{On the left, we see a 2D t-SNE projection of the vulnerability space (in fact, a random sample) using
          transformer-based embeddings color-coded by the year of publication. It is quite noticeable that the
          vulnerabilities from the same year are often clustered together (partly caused by temporal clustering of
          similar technologies and attacks). To look at a finer temporal resolution, on the right-hand side, we show the
          vulnerabilities for 2020 color-coded by day of the year. It may be less obvious,
          but there is a slight tendency for similar vulnerabilities to be published on the same day (mini-clusters),
          which can be often explained by related vulnerabilities arriving in small batches
          (e.g., because of a common underlying weakness).}
	\vspace{5pt}
	\label{fig-trans-year}
  \end{center}
\end{figure*}

\begin{figure*}[!htb]
  \begin{center}
	\includegraphics[width=\linewidth]{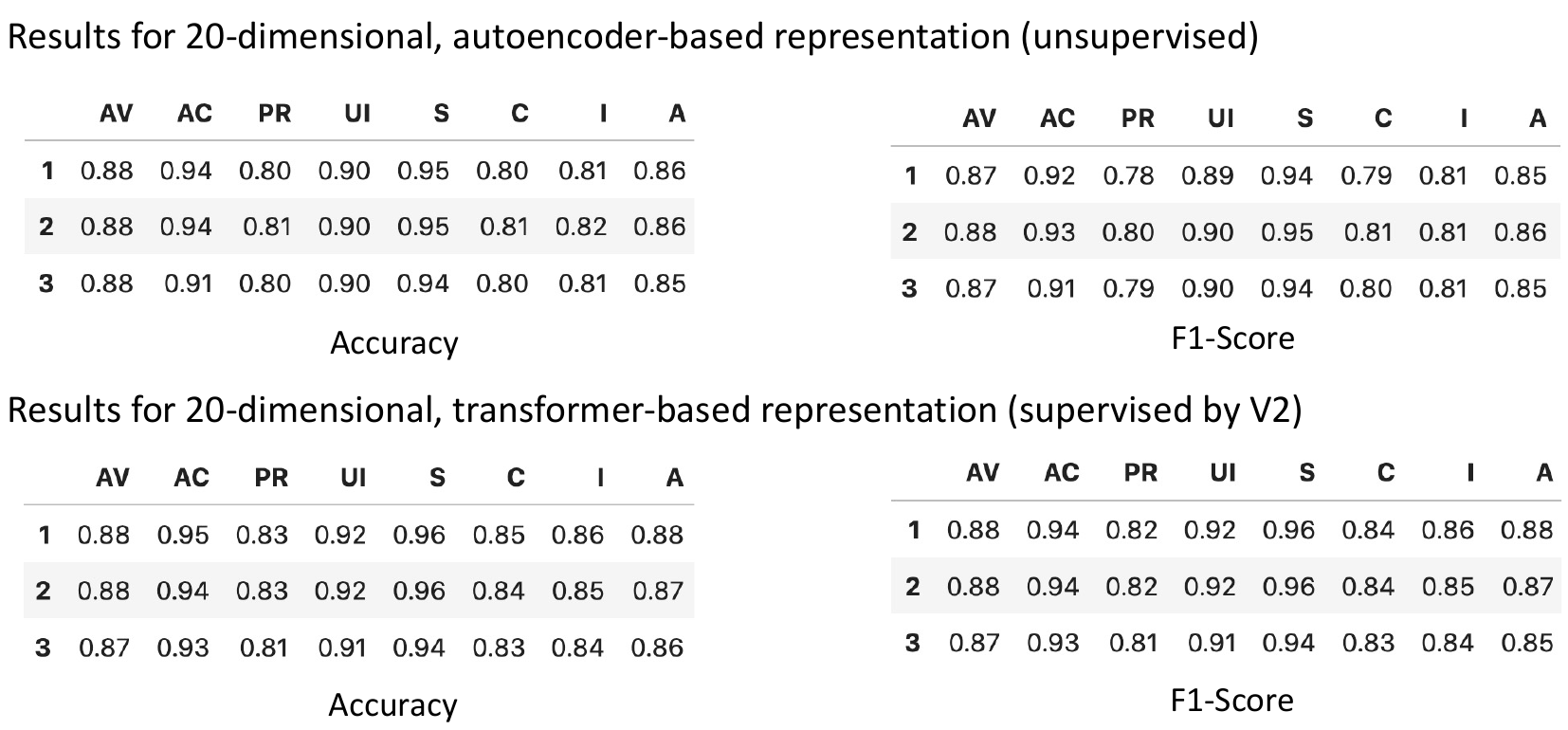}
	\caption{Sample Results for MLP Classifiers w.r.t. CVSS V3 Labels.
          Comparing two underlying models (autoencoder and transformer networks) the improvement in
          performance is quite notable in almost all labels. The CVSS V3 labels are:
          attack vector (AV), attack complexity (AC), privilege requirement (PR), user interaction (UI),
 	  change of scope (S), confidentiality (C), integrity (I), and availability(A). 
          The first column denotes the number of hidden layers, and we can see that
          one or two hidden layers are optimal (depending of the type of underlying model).}
	\vspace{5pt}
	\label{fig-classifier-eval}
  \end{center}
\end{figure*}

\begin{figure*}[!htb]
  \begin{center}
	\includegraphics[width=\linewidth]{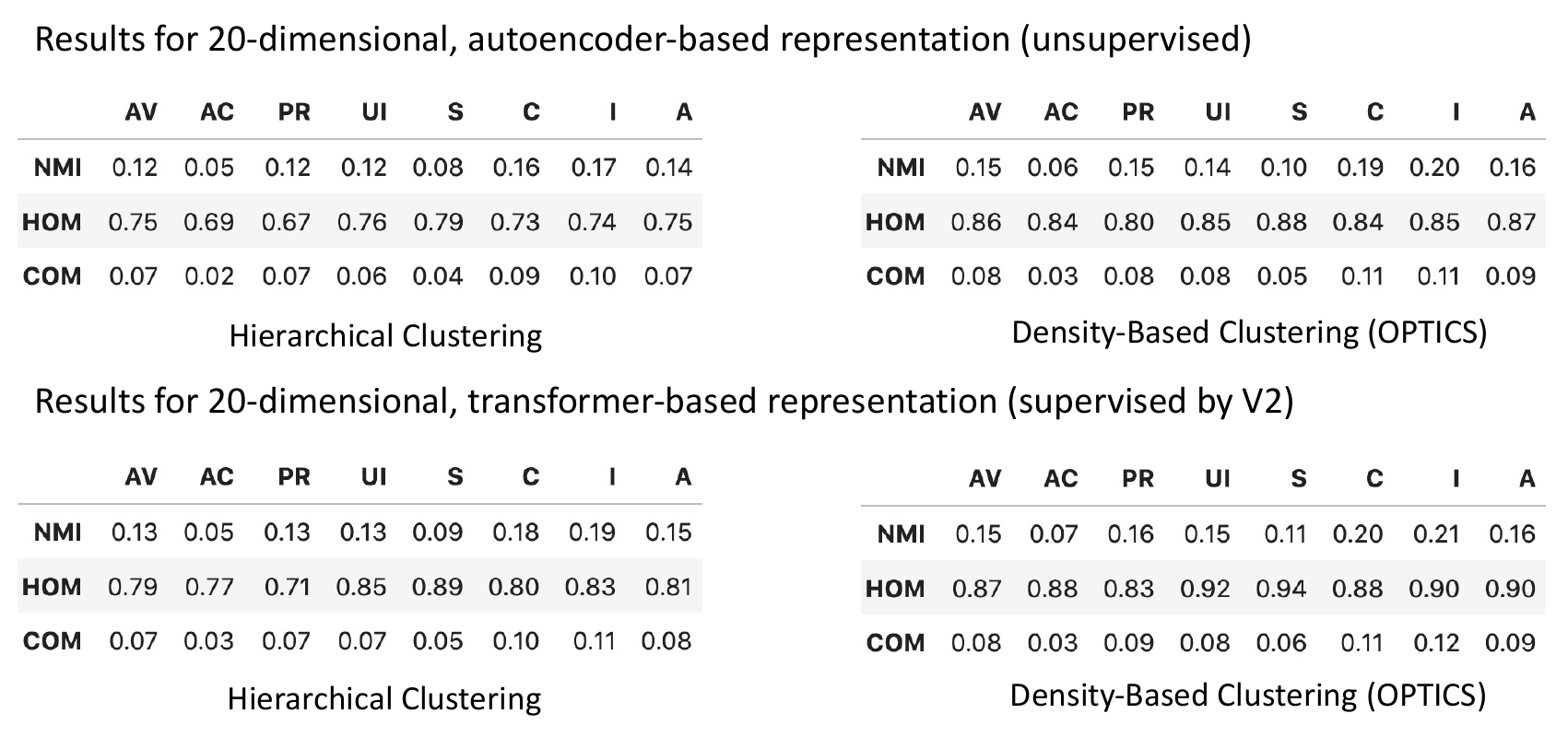}
	\caption{Sample Results for Clustering w.r.t. CVSS V3 Labels. 
          The first column in the tables denotes the information-theoretic evaluation measure.
          Comparing two underlying models (autoencoder and transformer networks), the improvement in
          performance again is quite notable in almost all labels (abbreviations as in Fig.~\ref{fig-classifier-eval}).
          Orthogonal improvements are possible due to the type of clustering algorithm. Here we compare hierarchical clustering
          and OPTICS as a representative of density-based clustering (which has the advantage of allowing partial clustering).}
	\vspace{5pt}
	\label{fig-clustering-eval}
  \end{center}
\end{figure*}

\clearpage

\section{Testing a Theory of Vulnerability Composition}

\begin{figure*}[!htb]
  \begin{center}
	\includegraphics[width=\linewidth]{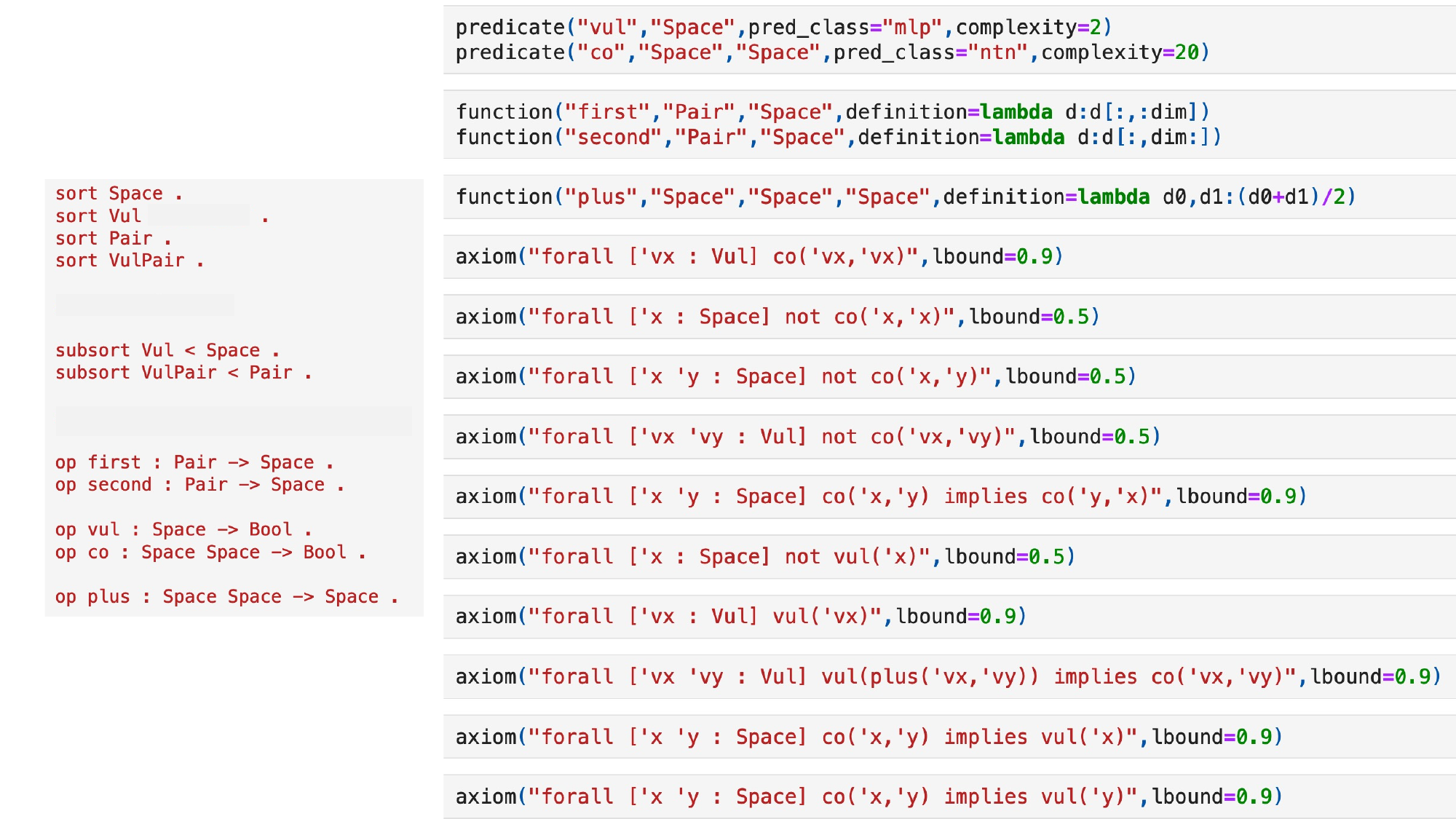}
	\caption{This figure shows the signature (left-hand side) and the axioms (right-hand side) of a simple
          test theory of compositionality used in our experiments with PALO/LIME. The space of potential vulnerabilities
          and known vulnerabilities are represented by sorts \texttt{Space} and \texttt{Vul}, respectively. Vulnerabilities
          that may exist (known or unknown) are (probabilistically) modeled as a learnable unary predicate (from the MLP family)
          and the binary composability relation \texttt{co} is a learnable relation (from the family of Neural Tensor Networks).
          A key axiom relates \texttt{co} and \texttt{plus}, which is defined as an average to approximate our NLP composition.
          All axioms are equipped with roughly estimated probabilistic lower bounds (to be concretized by the engine),
          as they clearly cannot be satisfied together in a classical sense. The models synthesized by
          our LIME engine approximately satisfy all these constraints (see Fig.~\ref{fig-theory-results}).
          Somewhat surprisingly, the best models in terms of respecting the axioms
          are those generated using a transfomer-based representation of vulnerabilities.}
	\vspace{5pt}
	\label{fig-theory}
  \end{center}
\end{figure*}

\begin{figure*}[!htb]
  \begin{center}
	\includegraphics[width=0.8\linewidth]{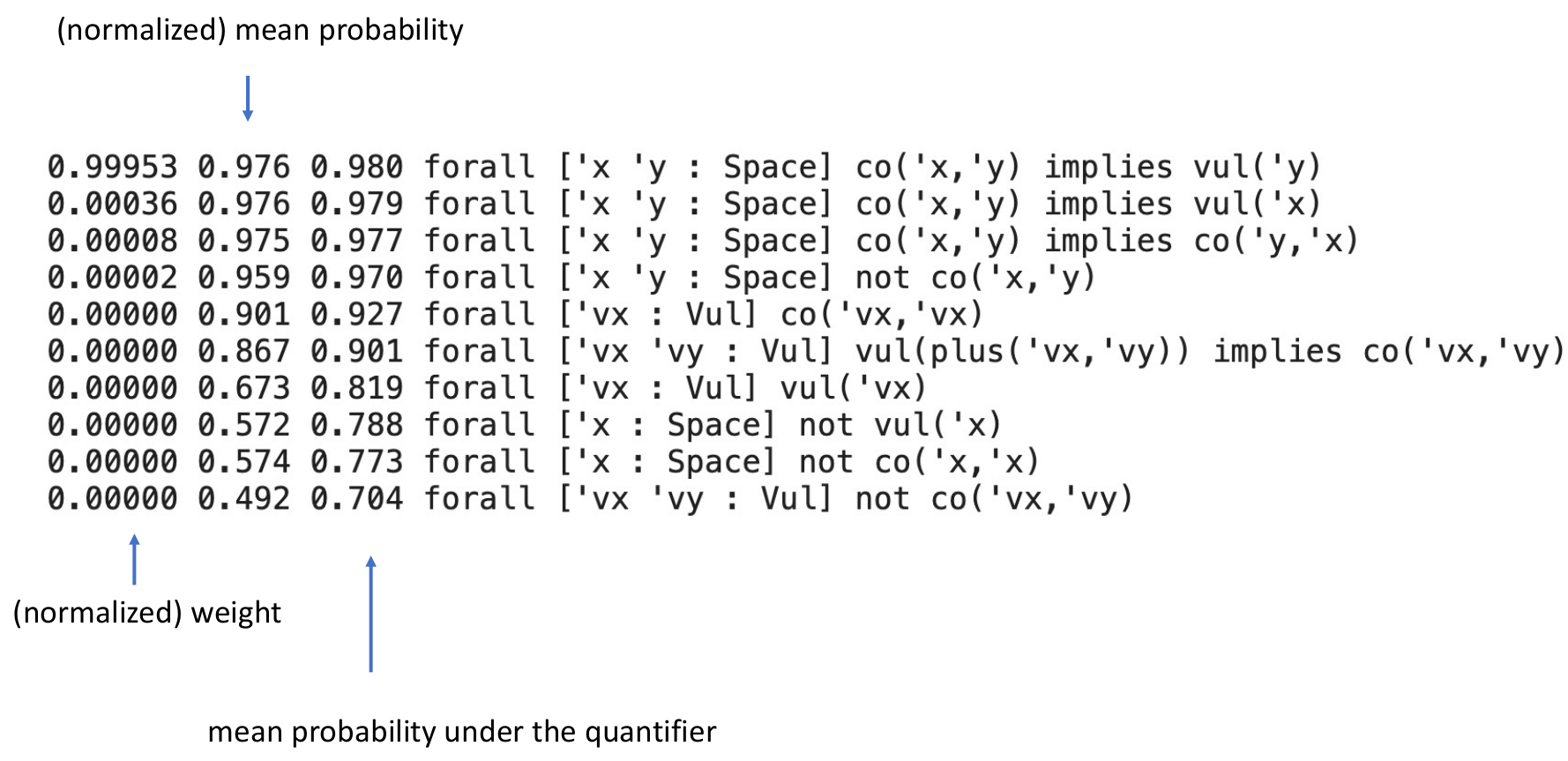}
	\caption{Sample model generated by LIME utilizing a transformer-based representation. The normalized
          mean probability is given for each axiom of the theory. Other models synthesized by the engine may
          lead to slightly different tradeoffs between the axioms. Hence, multiple models are generated
        in our parallel workflow.}
	\vspace{5pt}
	\label{fig-theory-results}
  \end{center}
\end{figure*}

\begin{figure*}[!htb]
  \begin{center}
	\includegraphics[width=0.8\linewidth]{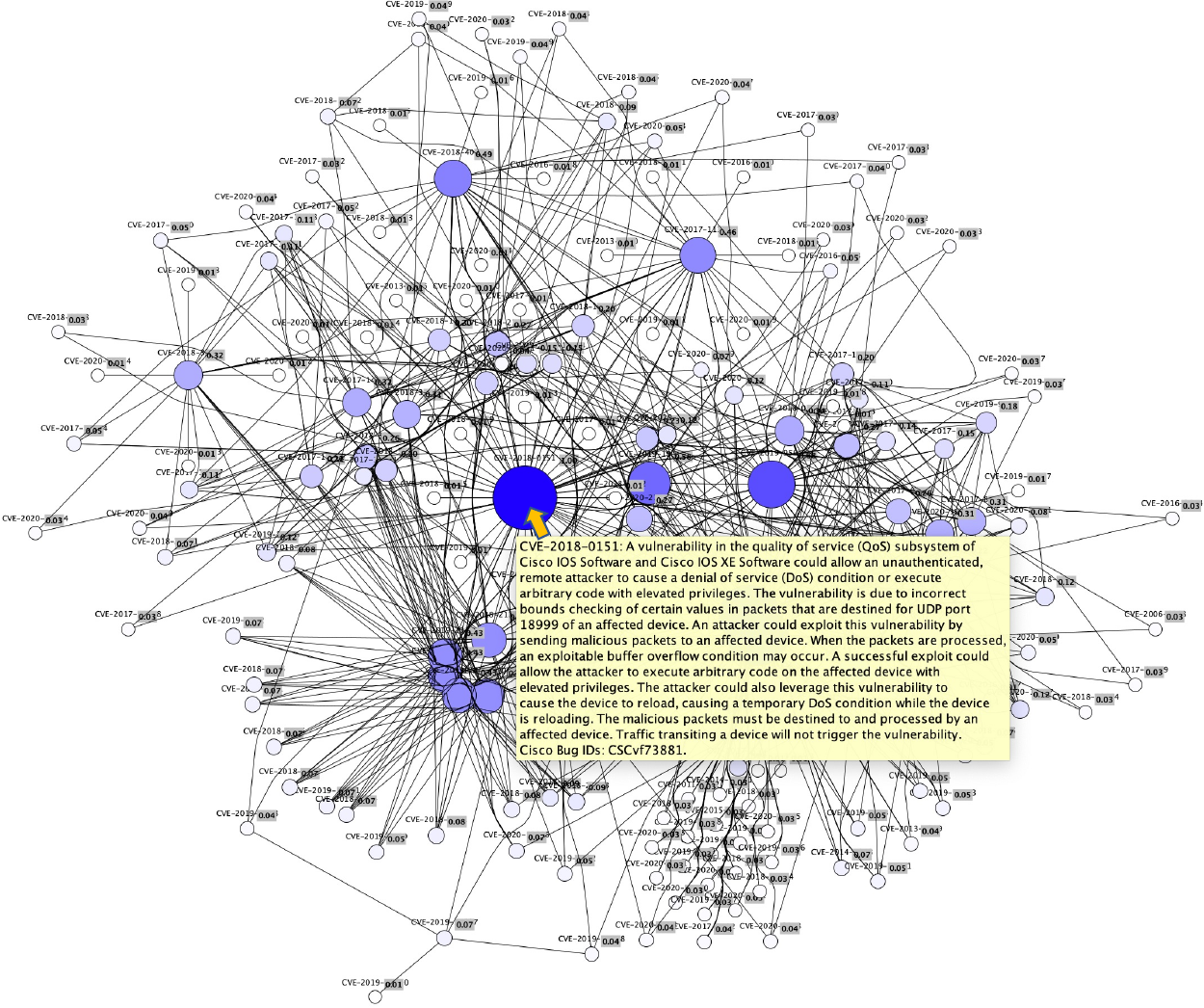}
	\caption{Partial graph of the composability relation (focussing on highest probability edges)
          associated with the model in Fig.~\ref{fig-theory-results}.
          The size of nodes highlights the number of connected edges (one possible centrality measure). In many cases,
          we found that nodes with high centrality are particularly suitable to be composed. In this figure, the
          highlighted node enables remote code execution, which can be potentially combined with many other vulnerabilities.}
	\vspace{5pt}
	\label{fig-theory-graph}
  \end{center}
\end{figure*}

\clearpage

\bibliographystyle{abbrv}
\bibliography{ref}

\end{document}